\documentclass{aa}  
\usepackage{graphicx,natbib}
\begin{document}
   \title{The extremely collimated bipolar H$_2$O jet from the
   NGC\,1333--IRAS\,4B protostar}


   \author{J.-F. Desmurs\inst{1} \and C. Codella\inst{2} \and J. Santiago-Garc\'{\i}a\inst{1,3}
          \and M. Tafalla\inst{1} \and R. Bachiller\inst{1}}

   \offprints{J.-F. Desmurs (desmurs@oan.es)}

\institute{Observatorio Astron\'omico Nacional (IGN), Alfonso XII 3, 
28014 Madrid, Spain \and
INAF, Istituto di Radioastronomia, Sezione di
Firenze, Largo E. Fermi 5, 50125 Firenze, Italy
\and
IRAM, Avenida Divina Pastora 7, 18012 Granada, Spain}

   \date{Received 18/11/2008 ; accepted 12/02/2009}


\titlerunning{The H$_2$O jet from the NGC\,1333--IRAS\,4B protostar}
\authorrunning{Desmurs et al.}

  \abstract 
{We have performed observations of water maser emission towards a
  sample of low-mass protostars, in order to investigate the properties
  of jets associated with the earliest stages of star formation and
  their interaction with the surrounding medium.}
{The main aim is to measure the absolute positions and proper motions
  of the H$_2$O spots in order to investigate the kinematics of the
  region from where the jet is launched.}
{We imaged the protostars in the nearby region NGC\,1333--IRAS\,4 in
  the water maser line at 22.2 GHz by using the VLBA in phase-reference
  mode at the milliarcsecond scale over four epochs, spaced by one month to
  measure proper motions.}
  {Two protostars (A2 and B) were detected in a highly variable H$_2$O
  maser emission, with an active phase shorter than four weeks. The
  H$_2$O maps allow us to trace the fast jet driven by the B protostar:
  we observed both the red- and blue-shifted lobes very close to the
  protostar, $\le$ 35 AU, moving away with projected velocities of
  $\sim$ 10--50 km s$^{-1}$. The comparison with the molecular outflow
  observed at larger scale suggests a jet precession with a 18$\arcmin$
  yr$^{-1}$ rate. By measuring the positional spread of the H$_2$O
  spots we estimate a jet width of $\sim$ 2 AU at a distance of $\sim$
  12 AU from the driving protostar.}
{}


   \maketitle
%

\section{Introduction}

The formation of a star is accompanied by a period of strong mass
ejection in which two lobes of supersonic gas move away from the newly
born object forming bipolar jets, strongly interacting with the natal
cloud.  The launching process of jets is still not well known, nor 
is the precise launch region, i.e. whether the jet originates
from the interface between the star's magnetosphere and disk or from a
wide range of disk radii \citep[e.g.][]{Shang07,Pudritz07}.  Most
investigations of the jet origin  so far have been made in atomic
microjets from optically visible T Tauri stars, thanks to the high
angular resolution available in the optical range
\citep{Ray07,Bally07}. As a next step, it is crucial to investigate the
properties of jets driven by protostars, the so-called Class 0 objects
\citep{Andre90}, which are still hidden in the innermost part of the
star forming region.  However, observations of these regions in the
optical and even in the IR spectral windows are not straightforward,
being hampered by the high attenuation associated with these high
density parts of the cloud.

In this context, water (H$_2$O) masers at 22 GHz represent a unique
tool to investigate the mechanism of jet formation and collimation in
the earliest star forming phases.  On the one hand, single-dish surveys
show for Class 0 objects a definitely higher detection rate than that 
found for the more evolved Class I ones
\citep[e.g.][]{Furuya03}.  On the other hand, images taken with the VLA
interferometer have confirmed a close association between water masers
and protostars, finding the H$_2$O spots concentrated within several
hundred AU of the central star \citep[e.g.][]{Terebey89, Chernin95,
Meehan98, Furuya05}.

However, there are only a few high angular resolution studies
available, using the Very Long Baseline Interferometry (VLBI), which is
needed to trace the innermost part of the jet/disk system at the
sub-AU-scales, in the closest vicinity of the exciting
object. \cite{Claussen98} and \cite{Furuya00} observed respectively the
HH212 and S106 FIR protostars, tracing with the H$_2$O maser spots the
curved bow shocks thought to travel along a jet. More recently,
\cite{Moscadelli06} reported multi-epoch VLBI observations of water
masers towards a sample of low-mass protostars in the Serpens region
and towards RNO 15-FIR.  For the Serpens Class 0 objects,
\cite{Moscadelli06} found high-velocity spots located perpendicular to
the jet axis: the authors suggest that such maser emission originates
at the base of the jet, tracing the interaction between the jet itself
and the accretion disk. For RNO 15-FIR, water masers are distributed
along a line, suggesting their association with a collimated flow, in
agreement with the results obtained by \cite{Marvel08} for the
NGC\,1333--IRAS\,4B region. Thus, the origin of H$_2$O masers in
low-mass protostars is far from being fully understood, calling for
further observations at AU-scales.

Here, we report the results of a multi-epoch H$_2$O maser survey
towards the NGC\,1333--IRAS\,4 cluster of protostars, performed with
0.28 AU resolution and in phase-reference mode.  Our main aim is
twofold: (i) to measure the absolute positions and proper motions of
the maser spots in order to study the kinematics of the disk/jet
systems, (ii) to provide new observations of low-mass protostars
allowing a step ahead in the discussion of the origin of jets.

\section{NGC\,1333--IRAS\,4}

NGC\,1333 is a well-known star forming region containing a large number
of protostars and located at a distance of 235 pc, according to the
recent observations \cite{Hirota08} performed with VERA (VLBI
Exploration of Radio Astrometry).  The IRAS\,4 region is split into
three star forming sites, called 4A, 4B, and 4C, which can be
identified in the continuum, from cm- to submm-wavelengths
\citep{Sandell91, Lefloch98, Rodriguez99}.  High angular resolutions
cm- and mm-observations \citep{Looney00, Reipurth02} have revealed that
IRAS\,4A is a binary system composed of A1 and A2 and is driving a
highly collimated outflow in the northeast-southwest direction
\citep{Choi05}. IRAS\,4B is also associated with a bipolar outflow
located in the North-South direction \citep{Choi01}. IRAS\,4C has been
revealed as a water maser source by single-dish observations
\citep{Haschick80}, although it was not detected by later
interferometric surveys \citep{Rodriguez02, Furuya03, Park07}.  All
these characteristics make IRAS\,4 an excellent laboratory in which to
study multiple star formation.

The IRAS\,4 star forming region has been observed in the 22 GHz H$_2$O
maser line by \cite{Park07} using the VLA and a resolution of $\sim$ 80
mas. These authors found no emission towards the A1, whereas they
observed six maser spots close to A2. Since the A2 maser group shows
velocities close to the systemic velocity, \cite{Park07} interpreted
such H$_2$O spots as a tracer of a circumstellar disk. On the other
hand, \cite{Marvel08} reported VLBA observations of H$_2$O masers
towards NGC\,1333--IRAS\,4B, finding two groups of maser spots
expanding away from each other at velocities of $\sim$78~km\,s$^{-1}$
along a direction different from that of the large scale outflow,
already traced with fews H$_2$O spots by \cite{Park07}. The
disagreement found by \cite{Marvel08} calls for further analysis in
order to clarify which are the driving sources and whether
NGC\,1333--IRAS\,4B has a companion as has been proposed by
\cite{Marvel08}.

\section{Observations}

\begin{table}
\caption[]{Coordinates used as phase centers at VLBA} 
\label{coordinates}
\begin{tabular}{ccc}
\hline
Source & $\alpha_{\rm J2000} $ & $\delta_{\rm J2000}$ \\
   & ($^{h}$ $^{m}$ $^{s}$) & ($\degr$ $\arcmin$ $\arcsec$) \\
\hline
IRAS\,4A1+A2  & 03:29:10.456 & +31:13:32.05 \\
IRAS\,4B      & 03:29:11.960 & +31:13:08.06 \\
IRAS\,4C      & 03:29:13.530 & +31:13:58.07 \\
\hline
\end{tabular}
\end{table}

\begin{figure*}
\includegraphics[angle=-90,width=16cm]{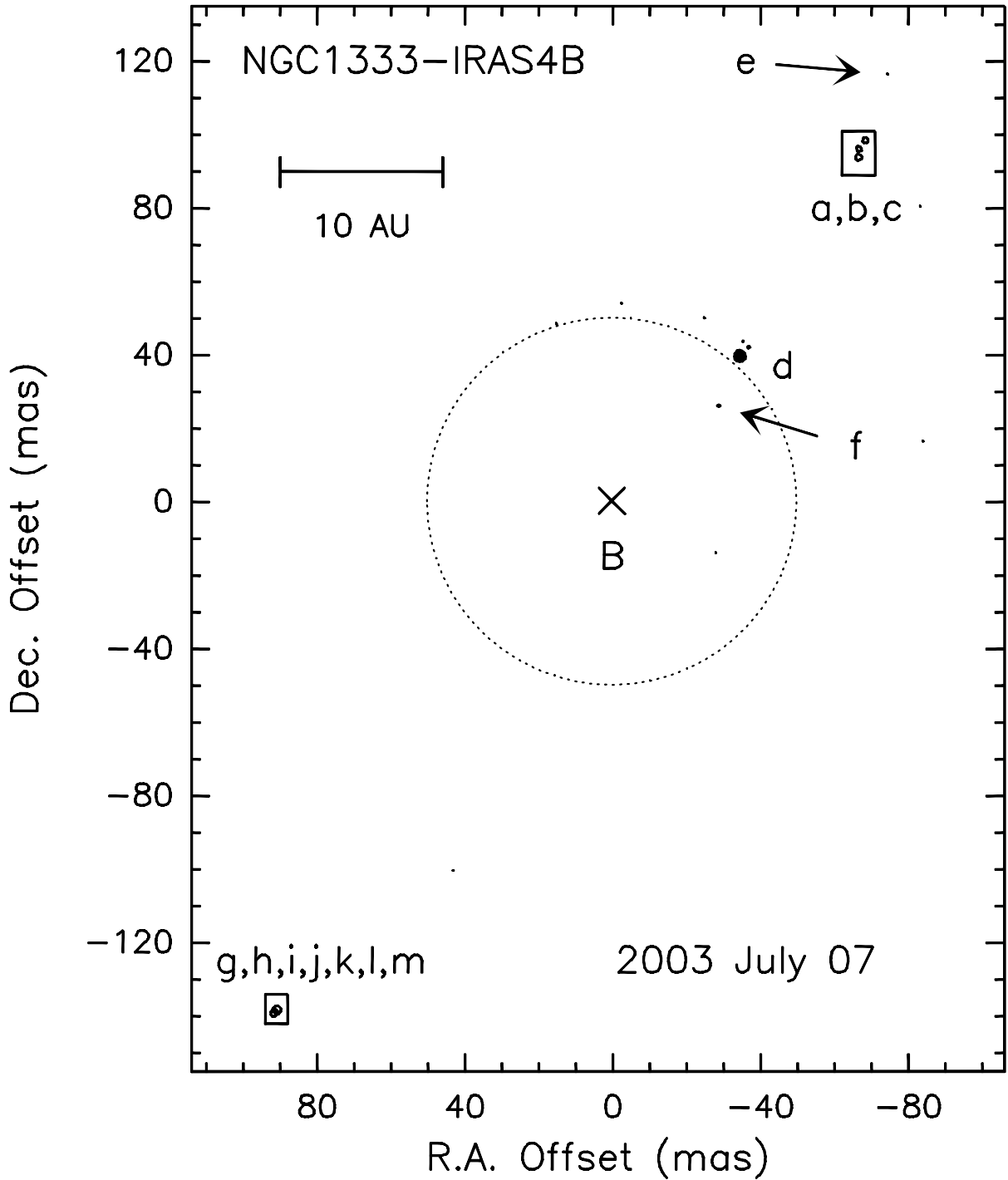}
\caption{Water maser map from Jul 07th observations showing a chain of
13 H$_2$O spots distributed over 73 AU, symmetrically located with
respect to the driving protostar IRAS\,4B.  The map is centered on the
coordinates of IRAS\,4B \citep[X-symbol;][]{Reipurth02}: $\alpha_{\rm
2000}$ = 03$^{\rm h}$ 29$^{\rm m}$ 12$\fs$003, $\delta_{\rm 2000}$ =
+31$\degr$ 13$\arcmin$ 08$\farcs$14.  The dotted circle indicates the
uncertainty (50 mas) associated with the position of the protostar. The
maser spots have been labelled with letters from `a' to `m'. For the
sake of clarity, only spots `d', `e' and `f' have been identified. Two
boxes identify the two regions where the rest of the spots have been
detected, as shown by the zoom-in of Fig. \ref{fig2} .  First contour
and step are $\sim$ 3 (0.84 Jy km s$^{-1}$) and 6$\sigma$,
respectively.}
\label{fig1}
\end{figure*}

\begin{figure*}
\includegraphics[angle=-90,width=16cm]{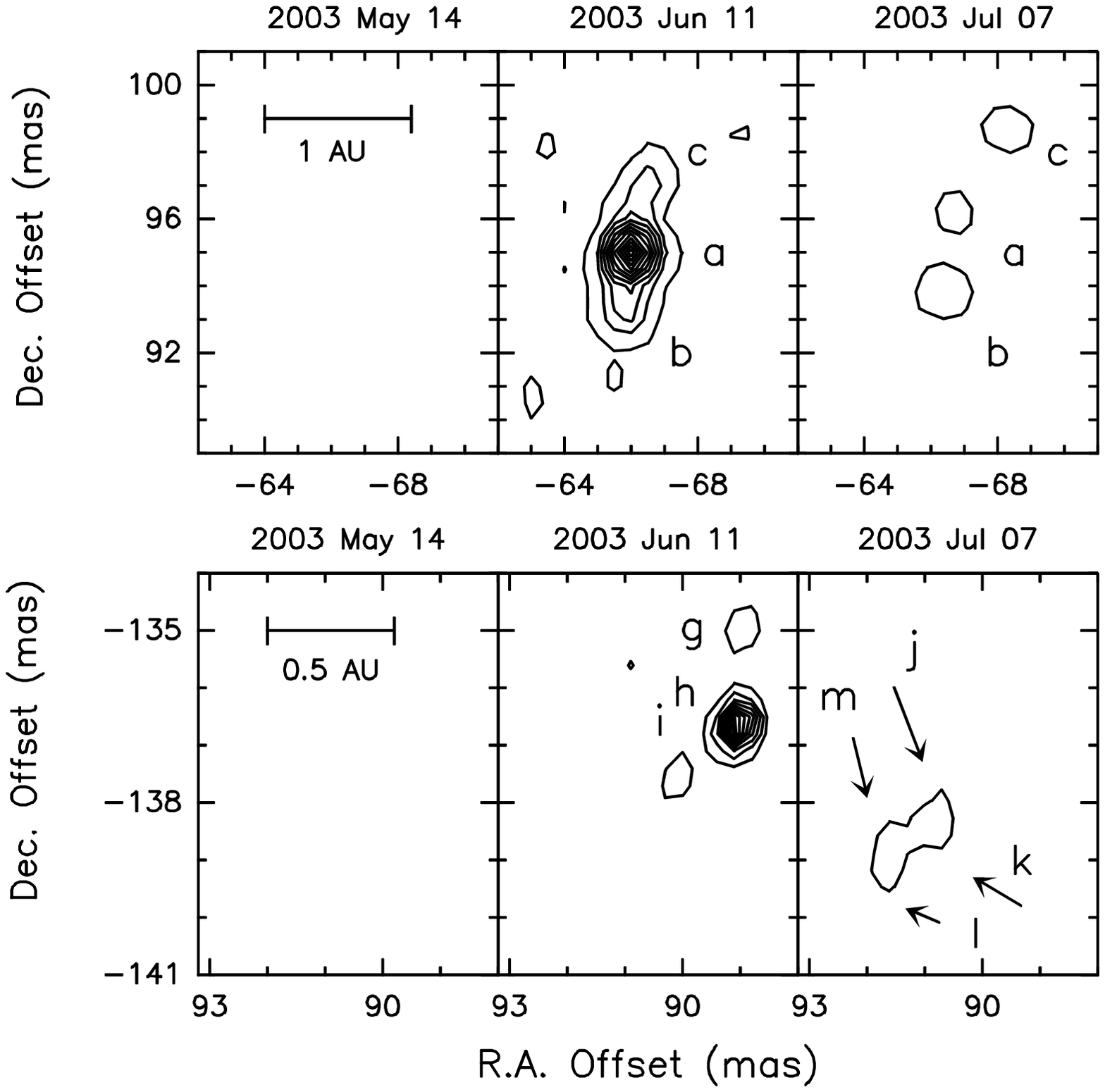}
\caption{Zoom-in of the two regions identified by black boxes in the
Jul 07th map of Fig. \ref{fig1} . The May 14th (no detection), June
11th, and July 07th epochs are reported.  The 1$\sigma$ level is 0.15
Jy km s$^{-1}$ for the June 11th map, and 0.28 Jy km s$^{-1}$ for the
July 07th map.  {\it Upper panels:} Northern region associated with the
red-shifted (see Fig. \ref{fig3} ) spots labelled `a', `b', and `c'.
First levels and steps are 3 and 12$\sigma$, respectively.  {\it Lower
panels:} Southern region associated with the blue-shifted (see
Fig. \ref{fig4} ) spots labelled `g', `h', `i', `j', `k', `l', and `m'.
First levels and steps are 3 and 6$\sigma$, respectively.  Note that
spots `i' and `j', as well as spots `k' and `l', are not well separated
in the present integrated H$_2$O emission map, but they are associated
with different velocities (see channel maps of Fig. \ref{fig5}).}
\label{fig2}
\end{figure*}
 
Using the VLBA network, we observed H$_2$O maser emission at 22.2 GHz
in phase-reference mode over four epochs spaced by about four weeks
during 2003 (Apr 1st, May 14th, June 11th, July 07th).  The duration of
each observation was 10 hours in total.

As the four protostars (IRAS\,4A1, 4A2, 4B, and 4C) fell in the primary
beam of the antenna, during the observation we tracked their geometric
barycenter.  As phase centers, we used the coordinates reported by
\cite{Rodriguez99} (Table \ref{coordinates}) in their VLA survey.  Data
were recorded in dual circular polarization with a velocity resolution
(i.e. channel width) of $\sim$0.1~km\,s$^{-1}$ and a total velocity
coverage of about 55~km\,s$^{-1}$.  To be able to measure and compare
the relative position of the water maser emission across the epochs, we
used the phase-reference technique using the close phase calibrator
0333+321 (1.5 degree away). For this, we adopted a cycle time of 1
minute which left us about 20 seconds of integration time on our
calibrator and our target including slewing time of the antennas, for a
total coherent integration time on our target sources of about 2.5 h.

Data were correlated at the VLBA correlator in Socorro (New
Mexico, USA) we made three separate passes (one per source), using the
coordinates of each three sources (IRAS\,4 A,B,C) with an integration
time of 2 seconds.  The data reduction was performed using the
Astronomical Image Processing System (AIPS) package. The determination
of the phase calibration was done following the standard method for
spectral line VLBI data. The amplitude was calibrated using the
template spectra method.  The calibrator used to calculate the bandpass
correction was 3C84, and the single-band delay corrections which was
later applied to the spectral line sources was derived from the ICRF
calibrator $0333+321$. The fringe rates were estimate by selecting a
bright and simple-structured channel and residual phase errors were
removed later in a self-calibration process. The corrections were
subsequently applied to the maser source to generate the channel
maps. The final restoring beam of 1.2 x 1.2 mas and each channel was
clean down to a final residual rms noise of about 15 mJy.

\section{Results and discussion} 

In the first epoch (Apr 1st), no emission was detected towards the four
sources. In particular, IRAS\,4C has never been detected in H$_2$O,
confirming the results from previous interferometric studies (see
Sect. 2).  The other protostars (IRAS\,4A1+4A2 and IRAS\,4B) presented
water emission in two of the four epochs observed. This confirms the
high variability of the H$_2$O spots associated with low-mass young
stellar objects \citep{Wilking94, Claussen96, Marvel08}, with an active
phase shorter than four weeks.  Table \ref{spots} summarises the
position, radial velocity, and flux of the observed H$_2$O spots.

\subsection{NGC\,1333-IRAS\,4B}

\begin{figure}
\includegraphics[angle=-90,width=8cm]{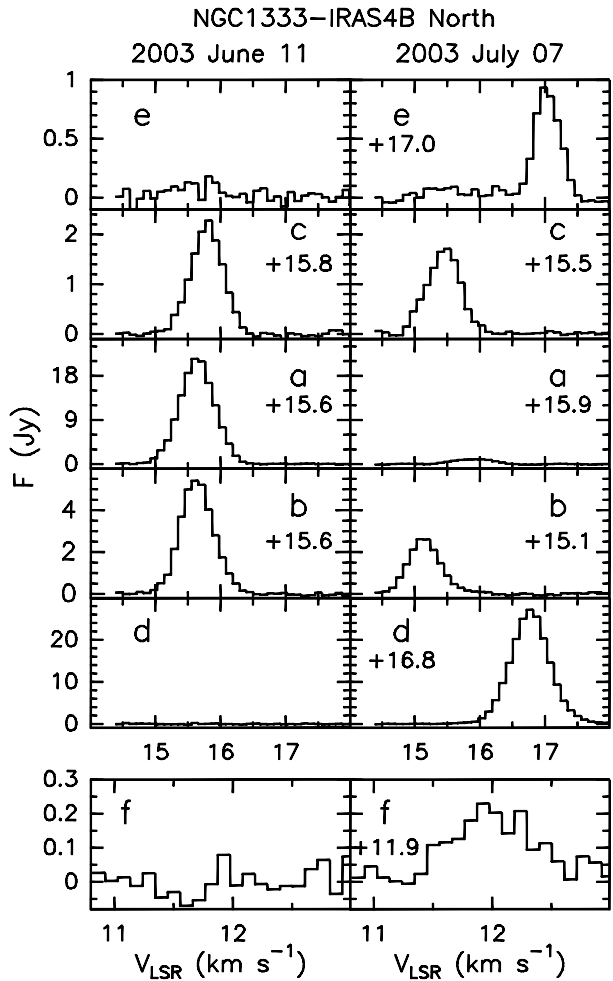}
\caption{Spectra of the water maser emission in IRAS\,4B associated
with the red-shifted northern spots and observed on June 11th and July
07th (see Figs. \ref{fig1} and \ref{fig2}).  The ambient LSR velocity
is +7.25 km s$^{-1}$ \citep{Mardones97}.  The peak velocity (km
s$^{-1}$) is indicated in each box.}
\label{fig3}
\end{figure}

\begin{figure}
\includegraphics[angle=-90,width=8cm]{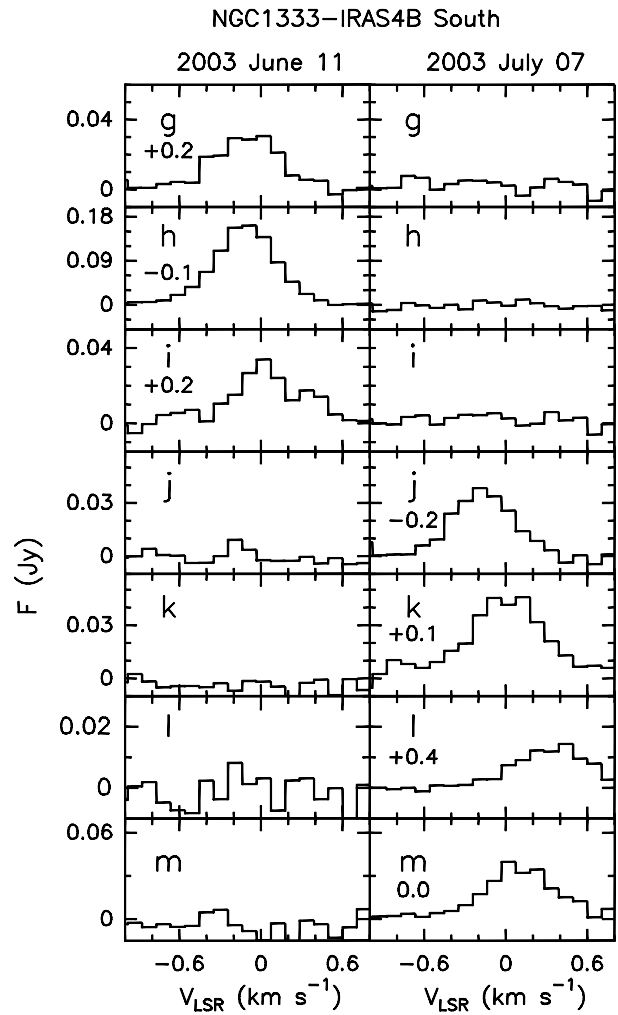}
\caption{Spectra of the water maser emission in IRAS\,4B associated
with the blue-shifted southern spots and observed on June 11th and July
07th (see Figs. \ref{fig1} and \ref{fig2} ).  The ambient LSR velocity
is +7.25 km s$^{-1}$ \citep{Mardones97}.  The peak velocity (km
s$^{-1}$) is indicated in each box.}
\label{fig4}
\end{figure}

\begin{figure*}
\includegraphics[angle=-90,width=16cm]{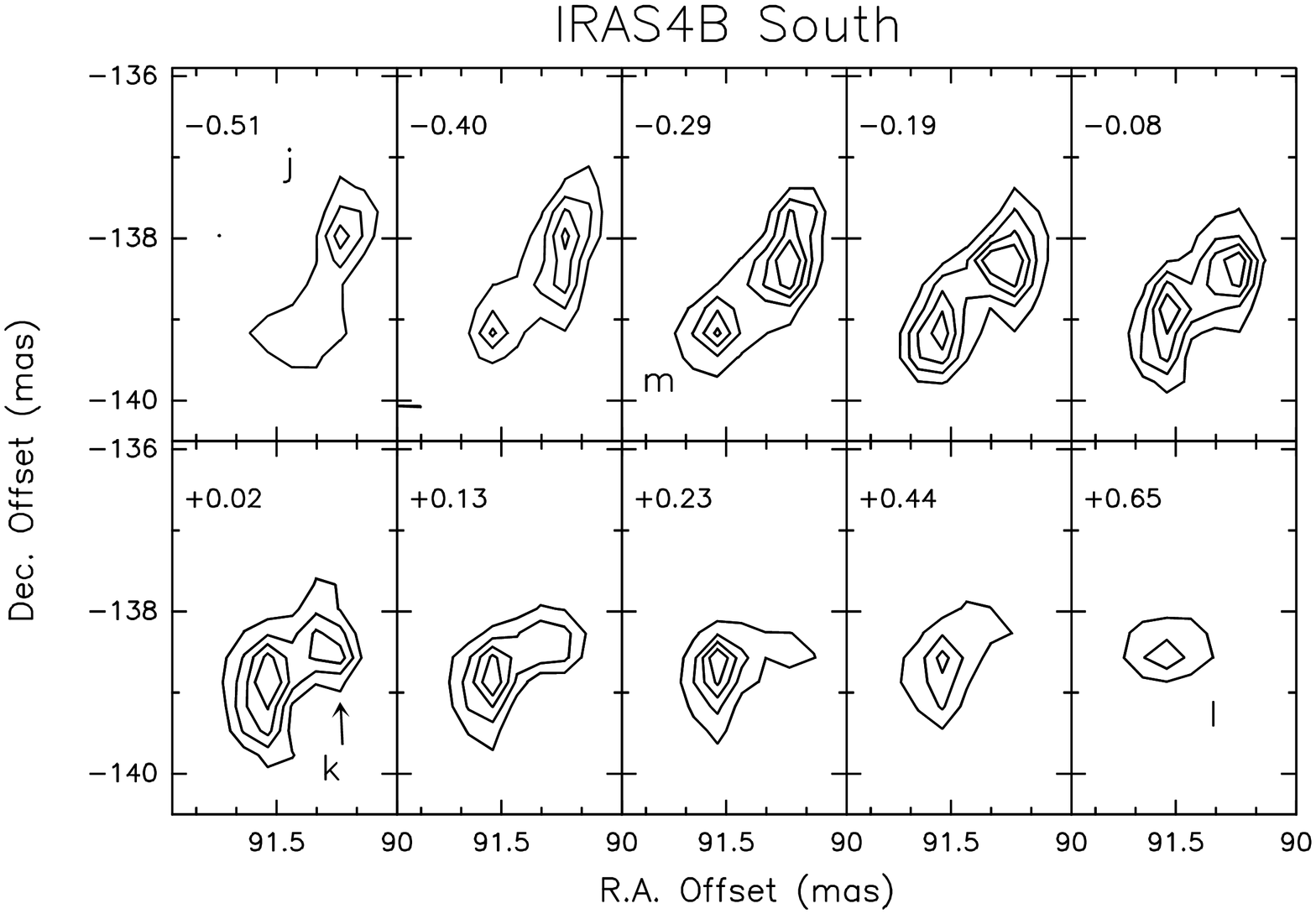}
\caption{Channel maps of the H$_2$O emission towards the IRAS\,4B
southern region, where the blue-shifted spots have been detected.  The
maps correspond to the 2003 July 07 epoch.  Each panel shows the
emission integrated over a velocity interval of $\sim$ 0.11 km s$^{-1}$
centered at the value given in the left corner.  The ambient velocity
emission is +7.25 km s$^{-1}$ \citep{Mardones97}.  Both first contours
and steps are 0.05 ($\sim$5$\sigma$) Jy km s$^{-1}$.  The labels
indicate the four H$_2$O spots (see also Fig. \ref{fig3} ).}
\label{fig5}
\end{figure*}

In IRAS\,4B water emission was detected on June 11th and July 07th.
Figure \ref{fig1} reports the July 07th map: up to 13 H$_2$O spots
(labelled `a' to `m') showing a remarkable degree of collimation are
seen.  As a reference, the cross in Fig. \ref{fig1} reports the
coordinates of B following \cite{Reipurth02} who gave the position of
the NGC\,1333 protostars, observed at 3.6 cm, with a 50 mas uncertainty
(dotted circle).  The spot `f' is the closest one to the B coordinates,
being inside the uncertainty area. Two boxes located at about (--70
mas,+100 mas) and (+70 mas,--130 mas) mark the regions where the
majority of the spots have been detected and that are shown in the
zoom-in of Fig. \ref{fig2}.  In particular, these spots (`a' to `c':
Upper panels; `g' to `m': Lower panels) are the only ones detected in
two epochs (Jun 11th and July 07th).  The high degree of variability of
the H$_2$O maser emission is clearly shown by the comparison between
the integrated emission maps of different epochs, as well as the H$_2$O
line profiles reported in Figs. \ref{fig3} and \ref{fig4} .  Note that
the present angular resolution does not allow us to disentangle spots
`i' and `j' as well as spots `k' and `l'.  However, these spots are
well identified by different radial velocities, as shown by the spectra
and by the channel maps of Fig. \ref{fig5}.

The NW water emission (spots `a' and `f') is red-shifted, with
velocities between +11.9 and +16.8 km s$^{-1}$ ($V_{\rm LSR}$ = +7.25
km s$^{-1}$; \cite{Mardones97}, whereas the SE emission (spots `g' and
`m') is blue-shifted (--0.1 to +0.4 km s$^{-1}$).  Thus, the positions
and velocities of the water spots suggest that they are tracing a jet
with a position angle of P.A. $\sim$ --32$\degr$ and driven by the 4B
protostar.  

\begin{figure*}
\includegraphics[angle=-90,width=18cm]{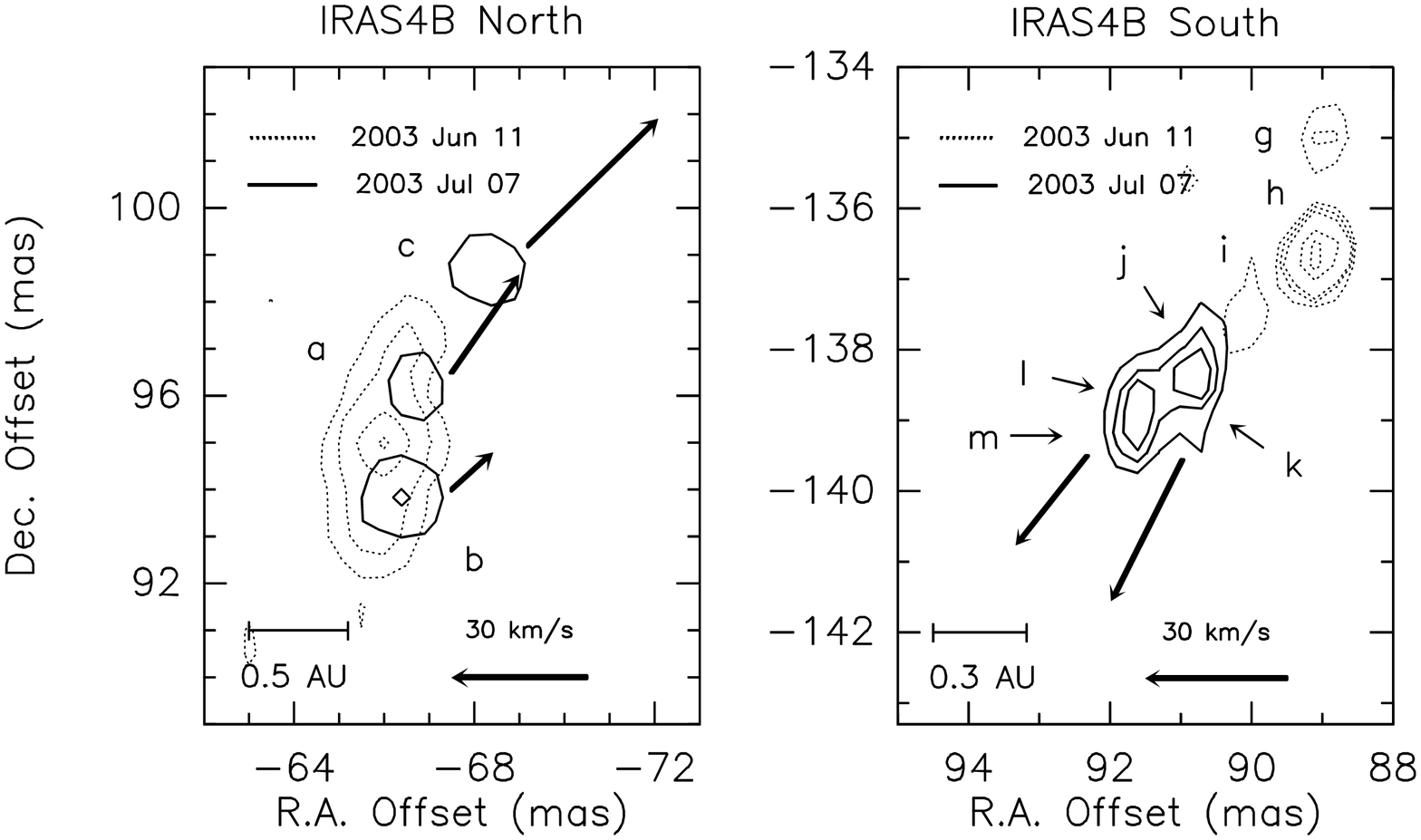}
\caption{Zoom-in of the northern (Left: spots `a', `b', and `c') and
southern (Right: spots `f' to `l') IRAS\,4B regions, where H$_2$O
emission has been detected.  The contour maps show the comparison
between the June 11th (dotted line) and July 07th (solid line) epochs.
For the sake of clarity only selected contour levels have been drawn.
Note that spots `i' and `j' as well as spots `k' and `l' are not fully
spatially resolved, but are associated with different velocities (see
spectra in Fig. \ref{fig4} and channel maps of Fig. \ref{fig5}).  Solid
black arrows indicate the apparent motions (parallax-subtracted): the
water spots are moving away from the protostar at projected velocities
of between 10 and 52 km s$^{-1}$.}
\label{fig6}
\end{figure*}

The pattern formed by the `a', `b', and `c' spots has been identified
in both Jun 11th and July 07th epochs.  Consequently, we have been able
to derive the apparent proper motion (see Fig. \ref{fig6}), once having taken
into account the parallax motion (including differential Galactic
rotation) of such emission. Water masers are moving towards the NW
direction with projected velocities in the range 10--40 km s$^{-1}$
(see black arrows).  The SE spots show different patterns in the two
epochs: we can therefore only arbitrarily assume they are tracing the
same components and then derive a typical proper motion: $\sim$ 40--50
km s$^{-1}$. Table \ref{motions} reports the derived proper motions.
These values are well in agreement with the results found by
\cite{Marvel08}, who found with their observations performed in 1998 a
total separation velocity of 78 km s$^{-1}$.  Note that for both the NW
and SW spots, the measured directions have a position angle between
--30$\degr$ and --45$\degr$, in agreement with the jet position angle,
again indicating a very collimated flow from the B protostar. By using
the measured proper motions and the radial velocity of the spots, we
derive a high inclination to the plane of the sky between 10$\degr$ and
35$\degr$, in agreement with the observed definite separation between
the red and blue lobes of the jet.

\begin{figure}
\includegraphics[angle=0,width=8cm]{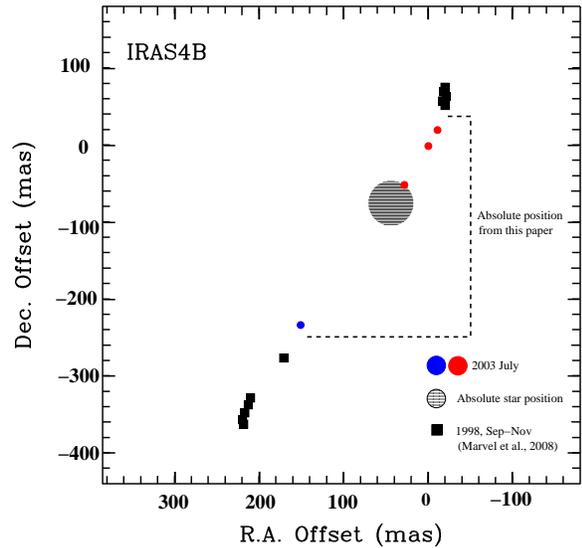}
\caption{Comparison between the position angle of the H$_2$O chain as
observed by \cite{Marvel08} without phase referencing (black squares)
and as reported in the present paper (2003 July; filled squares; red
and blue in the electronic edition only).  After 5 years, the position
angle remains the same (see text). The large circle is the position and
corresponding uncertainty of the B protostar.}
\label{fig7}
\end{figure}

Figure \ref{fig7} compares the position angle and the length of the
H$_2$O jet as observed in 1998 by \cite{Marvel08} and in 2003 by us.
Since the 1998 data are not in phase-reference mode we cannot
positionally overlap the two maps and thus the plot of Fig. \ref{fig7}
is just indicative.  However, it is possible to see that, after 5
years, the position angle, taking into account the corresponding
uncertainties, remains nearly the same (--29$\degr$ vs. --32$\degr$).
Interestingly, a chain of water maser spots with the same inclination
(--29$\degr$) was observed in 2006 by \cite{Park07}, see their Fig. 3
in their VLA survey.

The morphology and the kinematics of the H$_2$O spots can be compared
with the molecular outflow detected in several species (CO, CS, HCN,
CH$_3$OH, SiO, H$_2$CO) at larger scales (HPBWs between 1$\arcsec$ and
5$\arcsec$) by \cite{Choi01} and \cite{Jorgensen07} and thought to be
driven by IRAS\,4B. The molecular maps indicate a clear bipolar
structure, with a northern red-shifted lobe and a southern blue-shifted
lobe, as shown by the H$_2$O jet. The present maps suggest that, thanks
to the H$_2$O spots, we are tracing the jet sweeping up the surrounding
gas and creating the larger scale molecular outflow.  However, the
molecular outflow is located along the north-south direction, although
the blue lobe slightly bends to S-SE. Therefore its position angle
($\sim$ $0\degr$) is different from that of the H$_2$O chain. As the
inclination angle of 10-35 degrees excludes projection effects, either
the molecular outflow is driven by another jet, not detected in H$_2$O,
and driven by an unresolved companion of IRAS\,4B, or the difference
between the position angles is due to deflection or precession.  In the
first case, the jet could be bended by its interaction with the
surrounding high-density medium. However, this effect seems to be ruled
out for two main reasons: (i) the northern red-shifted and the southern
blue-shifted lobes are rotated by the same angle, and (ii) according to
C$^{18}$O and CS maps \citep{Choi01, Jorgensen07}, there is no evidence
of high-density clumps located along the main axis which could deflect
the jet direction.  On the other hand, if precession has occured the
jet should have changed about 30$\degr$ in around 110 years, which is
the dynamical scale of the molecular outflow, using the arcsec-scale
maps of \cite{Jorgensen07} and assuming a typical velocity of 100 km
s$^{-1}$.  In fact, the dynamical scale of the maser emission is of the
order of a year. In other words, we should conclude that the IRAS\,4B
is precessing with a definitely high rate: $\sim$ 18$\arcmin$
yr$^{-1}$.

\begin{table}
\begin{minipage}[t]{\columnwidth}
\caption[]{Offset position to the map center coordinates, radial
velocities and peak fluxes of the observed H$_2$O spots}
\label{spots}
\setlength{\tabcolsep}{3pt}
\begin{tabular}{cccccc}
\hline
Spot & epoch  & $\alpha_{\rm J2000} $ & $\delta_{\rm J2000}$ & $V_{\rm rad}$ & $F_{\rm peak}$$^a$ \\
     & (2003) & (mas) & (mas) & (km s$^{-1}$) & (mJy)\\
\hline
\multicolumn{6}{c}{IRAS 4B} \\
\hline
a    & Jun 11 & 66.0   & 95.0    & +15.6(0.1) & 21435(1850) \\
     & Jul 07 & 66.8   & 96.0    & +15.9(0.1) &  981(6)  \\
b    & Jun 11 & 65.8   & 93.2    & +15.6(0.1) &  5305(45)  \\
     & Jul 07 & 66.2   & 93.8    & +15.1(0.2) &  2603(19)  \\
c    & Jun 11 & 66.5   & 97.0    & +15.8(0.1) &  2375(110) \\
     & Jul 07 & 68.4   & 98.5    & +15.5(0.1) &  1709(12) \\
d    & Jun 11 & --     & --      & --         &  $\le$ 16 \\
     & Jul 07 & --33.0 & 40.0    & +16.8(0.1) & 25013(1202) \\
e    & Jun 11 & --     & --      & --         &  $\le$ 8 \\
     & Jul 07 & --75.0 & 120.0   & +17.0(0.1) &  850(59) \\
f    & Jun 11 & --     & --      & --         &  $\le$ 20 \\
     & Jul 07 & --28.0 & 26.0    & +11.9(0.2) &   230(25) \\
g    & Jun 11 & 89.0   & 135     &  +0.2(0.2) &   31(2)  \\
     & Jul 07 & --     & --      & --         &  $\le$ 5 \\
h    & Jun 11 & 89.1   & 136.6   & --0.1(0.1) &  164(5)  \\
     & Jul 07 & --     & --      & --         &  $\le$ 4 \\
i    & Jun 11 & 90.1   & 137.5   &  +0.2(0.1) &   40(2)  \\
     & Jul 07 & --     & --      & --         &  $\le$ 5 \\
j    & Jun 11 & --     & --      & --         &  $\le$ 6 \\
     & Jul 07 & 90.5   & --138.0 & --0.2(0.1) &   40(2)  \\
k    & Jun 11 & --     & --      & --         &  $\le$ 3 \\
     & Jul 07 & 90.7   & --138.5 &  +0.1(0.2) &   49(4)  \\
l    & Jun 11 & --     & --      & --         &  $\le$ 6 \\
     & Jul 07 & 91.55  & --138.5 &  +0.4(0.2) &   13(2)  \\
m    & Jun 11 & --     & --      &  --        &  $\le$ 3 \\
     & Jul 07 & 91.55  & --139.2 &  0.0(0.1)  &   31(3)  \\
\hline
\multicolumn{6}{c}{IRAS 4A2} \\
\hline
a    & May 14 & 00.0 & 00.0  &  +8.3(0.1) & 330455(16010) \\
     & Jun 11 & -0.2 & 2.0 &  +5.8(0.1) &  62545(4913) \\
b    & May 14 & --   & -- &  --        &  $\le$ 19 \\
     & Jun 11 & -3.1 & 7.0 &  +8.4(0.1) &  10512(990) \\
\hline
\end{tabular}
\begin{center}
$^a$ When H$_2$O emission is detected, we report here the peak flux and the corresponding
Gaussian fit uncertainty, otherwise we report the 1$\sigma$ r.m.s. as upper limit.  
\end{center}
\end{minipage}
\end{table}

\begin{table}
\caption[]{Derived proper motions of the IRAS 4B maser spots}
\label{motions}
\begin{tabular}{ccccc}
\hline
Spot & $\Delta  \alpha$ & $\Delta  \delta$ & $V_{\rm \alpha}$ & $V_{\rm \delta}$ \\
     & (mas) & (mas) & (km s$^{-1}$) & (km s$^{-1}$) \\
     \hline
a    & --0.96 &  +1.33 & 15.1 W &  20.9 N \\
b    & --0.59 &  +0.53 &  9.3 W &   0.8 N \\
c    & --1.86 &  +1.73 & 28.9 W &  27.1 N \\
h,k  &  +1.50 & --3.02 & 23.5 E &  47.0 S \\
i,m  &  +1.50 & --1.91 & 23.5 E &  28.9 S \\
\hline
\end{tabular}
\end{table}

\cite{Marvel08} noted that, in the case where the large scale outflow
was created 110 years ago by a clock-wise precessing jet observed now
in the water emission, the H$_2$O spots could be created by the impact
of the jet against the walls of the outflow cavities.  In this
scenario, the maser spots, which require extremely high-density
\citep[$\sim$ 10$^{7}$--10$^{9}$ cm$^{-3}$, e.g.][]
{Elitzur89,Kaufman96} conditions, could originate in shocked layers
compressed and accelerated at the interface with the surrounding
medium.  In particular, the maser amplification could be favoured at
the border of the cavity, where the optical path along the line of
sight is longer. This would also explain the high variability
associated with water masers. Actually, if we assume that maser
emission is produced in the thin layer at the border of the cavity, any
small change of geometry associated with these regions, expected to be
turbulent representing the interface between mass loss and molecular
cloud, would definitely change the optical path, thus affecting the
maser intensity.  This possible geometrical effect also would produce a
change in the projected velocity, as confirmed by the spectra reported
in Figs. \ref{fig3} and \ref{fig4}.  Interestingly, \cite{Marvel08}
predicted that further VLBA observations would reveal new masers south
(north) of the NW (SE) maser groups.  The comparison of our data with
those of \cite{Marvel08} presented in Fig. \ref{fig7} seems to support
this scenario, indicating for the H$_2$O map obtained in 2003 a
separation between the red- and blue-shifted emission smaller than that
observed by \cite{Marvel08} in 1998. Future phase referencing VLBA
observations would be instructive to check the validity of this
hypothesis.

Finally, it is tempting to give an estimate of the width of the H$_2$O
jet by using the angle defined by the positions of IRAS\,4B and spots
`e' and `f' (16$\degr$), and then measuring the width of this cone at
the border of the IRAS\,4B uncertainty area.  In this way we obtain a
width of 2 AU at a distance of $\sim$ 12 AU from the driving protostar.
Clearly, we cannot exclude that we are measuring the width of a cavity
swept-up by the jet itself. However, as far as we know, this is the
first attempt to estimate a Class 0 jet width so close to the
protostar.  Another estimate has been inferred for the HH212 jet by
using the H$_2$O spots imaged by \cite{Claussen98}.  In HH212 the width
is 18 AU at 50 AU from the driving source, indicating a decrease of the
jet collimation moving away from the driving protostar.  If we compare
the IRAS\,4B jet width with the range spanned by atomic microjets from
more evolved Class II sources surrounded only by thin disks \citep[see
Fig. 2 of] []{Cabrit07}, there is no evidence of a higher jet
collimation in Class 0 sources embedded in very dense cores.  This
confirms the conclusion reported by \cite{Cabrit07}, who concluded that
collimation is not due to external pressure, requiring that jets from
young stellar objects are self-collimated by internal magnetic
stresses.

\subsection{NGC\,1333-IRAS\,4A1+A2}

\begin{figure*}
\includegraphics[angle=-90,width=\textwidth]{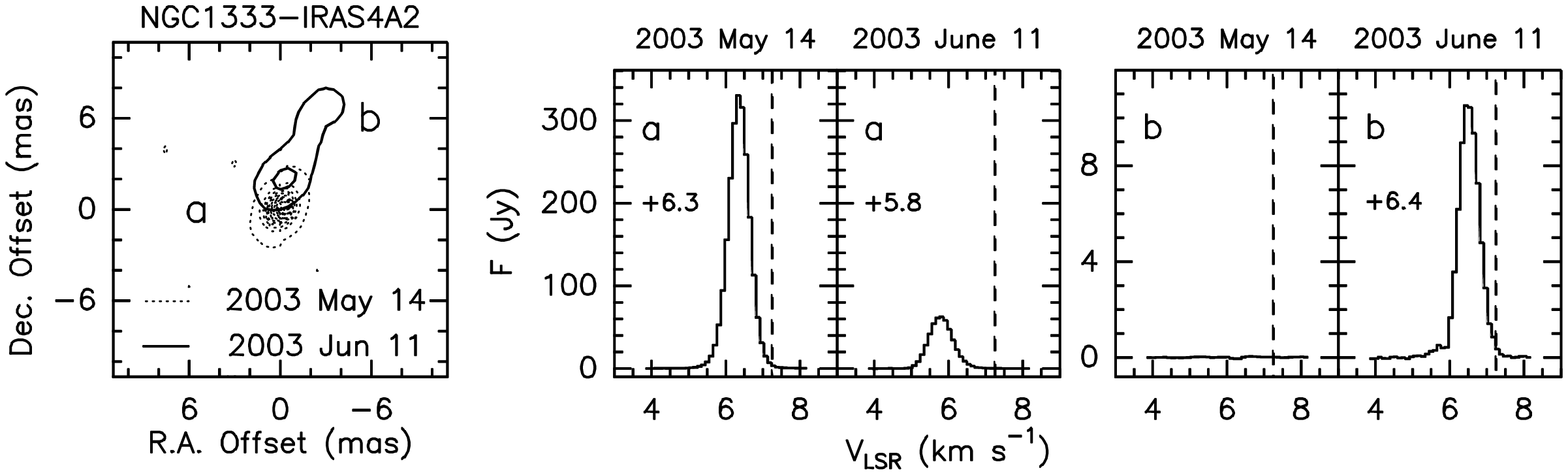}
\caption{{\it Left panel:} Water maser maps measured for IRAS\,4A
(dotted contour: May 14th; solid contour: Jun 11th).  First contour and
step are $\sim$ 3 (1.5 Jy km s$^{-1}$) and 13$\sigma$, respectively.
The A2 protostar ($\alpha_{\rm 2000}$ = 03$^{\rm h}$ 29$^{\rm m}$
10$\fs$421, $\delta_{\rm 2000}$ = +31$\degr$ 13$\arcmin$ 32$\farcs$21)
is located at +16,--74 mas, whereas the A1 protostar is at +1405,
--1235 mas, according to \cite{Reipurth02}.  {\it Right panels:}
Spectra of the H$_2$O maser spots: the peak velocity (km s$^{-1}$) is
reported in each box.  The vertical dashed line is for the ambient LSR
velocity \citep[+7.25 km s$^{-1}$; ][]{Mardones97}.}
\label{fig8}
\end{figure*}

In IRAS\,4A, maser emission was detected in the second and third epochs
(May 14th, Jun 11th). Figure \ref{fig8} (Left panel) summarizes the
obtained maps: only two spots (named `a' and `b') were detected. The
maps are centered on the coordinates of spot a as detected in May 14th:
$\alpha_{\rm 2000}$ = 03$^{\rm h}$ 29$^{\rm m}$ 10$\fs$4197221,
$\delta_{\rm 2000}$ = +31$\degr$ 13$\arcmin$ 32$\farcs$286030.  The
velocities of the spots lie in the +5.8,+6.4 km s$^{-1}$ range, thus
being slightly blue-shifted with respect to the ambient velocity
\citep[+7.25 km s$^{-1}$, ][]{Mardones97}.  Figure \ref{fig8} (Right
panel) reports the highly variable H$_2$O spectra.  Although the
positions of the protostars are known with higher uncertainty (50 mas)
than the H$_2$O spots (1 mas), the comparison between these positions
shows that the maser spots are very close to the A2 protostar, being
located at about 17 AU in the northern direction (offset: +16 mas, --74
mas).  On the other hand, A1 is definitely more distant from the maser
spots, at about 430 AU in the SE direction (+1405 mas, --1235
mas). This confirms the association of maser emission with A2, as also
found by \cite{Park07}. These authors interpreted their VLA maps as
mainly a sign of a circumstellar disk with, in addition, a few spots
related to a jet (see their Fig. 2).  Unfortunately, despite the high
resolution of the present data, since we have detected only two spots
we cannot infer any information about the morphology and the kinematics
of the masering gas or discuss if the A2 water masers are tracing
either a disk, a jet or a wide-angle cavity, as discussed for the
IRAS\,4B case.

\section{Summary}

We imaged the group of protostars in the NGC\,1333 IRAS\,4 region in
the H$_2$O maser line at 22.2 GHz by using the VLBA in phase-reference
mode.  The maps allow us to detect a chain of water maser spots tracing
the fast ($\ge$ 50 km s$^{-1}$) jet driven by the IRAS\,4B protostar.
We observed both the blue- and the red-shifted lobes very close to
IRAS\,4B, $\le$ 35 AU.  The H$_2$O chain is extremely collimated
suggesting a jet width of $\sim$ 2 AU.  The comparison with the large
scale molecular outflow suggests the jet is precessing with a rate of
18$\arcmin$ yr$^{-1}$.

\begin{acknowledgements}
We thank C. Goddi and L. Moscadelli for helpful discussions and suggestions.
\end{acknowledgements}

\bibliographystyle{aa}
\bibliography{1365}

\end{document}